\documentclass[%
 reprint,
 amsmath,amssymb,
 aps, prd, twocolumn, superscriptaddress
]{revtex4}

\usepackage{graphicx}
\usepackage{dcolumn}
\usepackage{bm}
\usepackage{amsmath}
\usepackage{amssymb}
\usepackage{hyperref}
\usepackage{color}

\newcommand{\refeq}[1]{Eq.~(\ref{eq:#1})}
\newcommand{\refeqs}[2]{Eqs.~(\ref{eq:#1})--(\ref{eq:#2})}
\newcommand{\refEq}[1]{Eq.~(\ref{eq:#1})}
\newcommand{\refEqs}[2]{Eqs.~(\ref{eq:#1})--(\ref{eq:#2})}
\newcommand{\reffig}[1]{Fig.~\ref{fig:#1}}

\newcommand{\refFig}[1]{Fig.~\ref{fig:#1}}
\newcommand{\refsec}[1]{Sec.~\ref{sec:#1}}
\newcommand{\refSec}[1]{Sec.~\ref{sec:#1}}

\begin{document}


\title{Constraining neutrino mass and dark energy with peculiar velocities and lensing dispersions of Type Ia supernovae}

\author{Aniket Agrawal}\email{aagrawal@asiaa.sinica.edu.tw}
\affiliation{%
Institute of Astronomy and Astrophysics, Academia Sinica, 11F of AS/NTU Astronomy-Mathematics Building, No.1, Sec. 4, Roosevelt Rd, Taipei 10617, Taiwan, R.O.C. 
}%

\author{Teppei Okumura}%
\affiliation{%
Institute of Astronomy and Astrophysics, Academia Sinica, 11F of AS/NTU Astronomy-Mathematics Building, No.1, Sec. 4, Roosevelt Rd, Taipei 10617, Taiwan, R.O.C. 
}
\affiliation{Kavli Institute for the Physics and Mathematics of the Universe (WPI), UTIAS, The University of Tokyo, Kashiwa, Chiba 277-8583, Japan}

\author{Toshifumi Futamase}
\affiliation{Department of Astrophysics and Meteorology, Kyoto Sangyo University, Kita-ku, Kyoto 603-8555, Japan}

\date{\today}

\begin{abstract}
We show that peculiar velocities of Type Ia supernovae can be used to derive constraints on the sum of neutrino masses, $\Sigma m_{\nu}$, and dark energy equation of state, $w = w_0+w_a(1-a)$, from measurements of the magnitude-redshift relation, complementary to galaxy redshift and weak lensing surveys. Light from a supernova propagates through a perturbed Universe so the luminosity distance is modified from its homogeneous prediction. This modification is proportional to the matter density fluctuation and its time derivative due to gravitational lensing and peculiar velocity respectively. At low redshifts, the peculiar velocity signal dominates while at high redshifts lensing {does}. We show that using lensing and peculiar velocity of supernovae from the upcoming surveys WFIRST and ZTF, {without other observations}, we can constrain $\Sigma m_{\nu} \lesssim 0.31$ eV, $\sigma(w_0) \lesssim 0.02$, and ${\sigma(w_a)} \lesssim 0.18$ {($1-\sigma$ CL)} in the $\Sigma m_{\nu}$-$w_0$-$w_a$ parameter space, where all the other cosmological parameters are fixed. {We find} that adding peculiar velocity information from low redshifts shrinks the volume of the parameter ellipsoid in this space by $\sim 33$\%. We also {allow $\Omega_{\text{CDM}}$ to vary as well as $\Sigma m_{\nu}$, $w_0$ and $w_a$, and demonstrate} how these constraints degrade as a consequence. 
\end{abstract}

\maketitle


\section{\label{sec:intro}Introduction}
The detection of non-zero neutrino masses~\cite{Fukuda:1998mi,Ahmad:2001an} provides conclusive evidence that the standard model of particle physics is incomplete. The importance of this discovery can be judged from the fact that two Nobel prizes have been awarded for work leading to this conclusion. Understanding the origin of their mass remains one of the key open questions in modern physics (see for e.g.~\cite{barger2012physics}). Cosmology can shed light on this problem through the dependence of the matter power spectrum~\cite{Takada:2005si} and growth rate of density fluctuations~\cite{Boyle:2017lzt} on neutrino masses. Another open problem is the nature of dark energy. Since its discovery in 1998~\cite{Perlmutter:1998np,Riess:1998cb} there has been a lot of effort in trying to explain what its nature is (for an observational overview see for e.g.~\cite{Weinberg:2012es}, and~\cite{Martin:2012bt} for issues on the theoretical side) but with limited success. In particular the $\Lambda$CDM model that is considered the ``standard" model of cosmology~\cite{ade:2015,Hinshaw:2012aka} fails to explain the observed value of $\Lambda$, the cosmological constant (see for e.g.~\cite{Martin:2012bt}). One interesting possibility is that it is driven by a scalar field~\cite{ratra:1998}, in which case its equation of state must vary with time. Upcoming surveys such as the Subaru Prime Focus Spectrograph (PFS)~\cite{Takada:2014,Tamura:2016wsg}, Dark Energy Spectroscopic Instrument (DESI)~\cite{DESI-Collaboration:2016}, Large Synoptic Survey Telescope (LSST)~\cite{abell2009lsst}, Euclid~\cite{Laureijs:2011gra} and others will pursue a stringent constraint on the sum of neutrino masses and time variation of dark energy equation of state as primary science goals. In these surveys, galaxy clustering and weak lensing will be primarily employed to achieve it. However, recent studies pointed out that Type Ia supernovae can also be used to determine the sum of neutrino masses, complementary to galaxy redshift and weak lensing surveys~\cite{Hada:2016dje,Hada:2018ybu}.

Type Ia supernovae are known to be \emph{standard candles}~\cite{hamuy:1996,riess:1996}. Their absolute luminosity can be accurately calibrated and thus one can use their apparent brightness to estimate distances to them, the so-called luminosity distances. The luminosity distance is a function of the background energy densities and the rate of expansion of the Universe, as described in~\refsec{magnitude}, thus offering a probe of these quantities. However, this dependence is limited to a homogeneous Universe. In the presence of inhomogeneities, this relation between the background quantities and luminosity distance is altered~\cite{Sasaki:1987ad} and the luminosity distance becomes a function of the degree to which the Universe deviates from the homogeneous one. This can be captured by the matter power spectrum and its time derivative, because the shape of the power spectrum is altered at wavelengths smaller than the free-streaming scale of neutrinos~\cite{Takada:2005si,Lesgourgues:2006nd,Boyle:2017lzt}. Since deviations from the homogeneous prediction are measured for individual supernovae, the key advantage is that one does not need to worry about issues of {complicated modeling of} bias and redshift space distortion that plague constraints from galaxy redshift surveys, or of intrinsic alignments and shape noise that plague weak lensing surveys. {Although} supernova surveys come with their own set of systematics, these are different from the above and so can provide complementary constraints on neutrino mass. 

Traditionally Type Ia supernovae have only been observed at low redshifts. For example, using the Hubble Space Telescope, a few Type Ia supernovae have been observed out to $z \sim 1.4$~\cite{Riess:2016jrr}, but the number of such supernovae is extremely low for use as cosmological probes of the perturbed Universe. Upcoming surveys such as Wide Field InfraRed Survey Telescope (WFIRST)~\cite{Spergel:2015sza}, however, will observe thousands of Type Ia supernovae out to $z \lesssim 1.7$ allowing us to gain significant insight into neutrino mass and dark energy using supernovae. In addition, surveys such as the Zwicky Transient Facility (ZTF)~\cite{Graham:2019qsw} will observe $\sim 2000$ supernovae out to redshift $0.1$ allowing one to probe neutrino mass and dark energy using a large sample of low redshift supernovae too. Previous studies of using the magnitude scatter to derive cosmological constraints have focussed on using either lensing or peculiar velocities. For instance, peculiar velocities of Type Ia supernovae have been used previously to derive constraints on growth of structure (see~\cite{Haugbolle:2007,Bhattacharya:2011,Turnbull:2012,Huterer:2017} and references therein). Here, we show for the first time that using peculiar velocities in addition to lensing can significantly improve not only the constraints on sum of neutrino masses but also those on the time variation of dark energy equation of state. One key point to note is that the redshifts considered here are of cosmological origin, but the total observed redshift of a given supernova  includes a component from its local velocity in the host galaxy. As such, the redshift of the host galaxy is additionally measured to obtain the peculiar velocity of cosmological origin, and that is the one we consider in the rest of the paper. In case the galaxy is a member of a cluster, the cluster redshift needs to be used, as galaxy peculiar velocities are also significantly affected by local velocities in a cluster~\cite{Leget:2018juj}. 

The rest of the paper is organised as follows. \refSec{magnitude} describes the observed scatter in brightness (magnitude) of supernovae coming from their peculiar velocities and lensing along the line of sight (l.o.s.), both of which are sourced by perturbations in the Universe. In~\refsec{neutrino} we discuss how non-zero neutrino masses affect this scatter. Finally, in~\refsec{results} we present the forecasts on neutrino mass from the two surveys, described in~\refsec{survey}, using the Fisher matrix formalism,~\refsec{fisher}. We conclude in~\refsec{conclusion}. 

\section{\label{sec:magnitude}The magnitude-redshift relation}

The starting point for extracting cosmological information from observations of Type Ia Supernovae is the magnitude-luminosity distance relation,
\begin{equation}\label{eq:m_dL_eq}
    m_{\text{obs}}(z) = 5\, \text{log}_{10} d_{\text{L}}(z)+M \,,
\end{equation}
where $M$ is the absolute magnitude of a supernova (including all corrections such as dust and reddening). {The function $d_{\text{L}}(z)$ is the luminosity distance, given by} 
\begin{equation}\label{eq:dL_z}
    d_{\text{L}}(z) = (1+z)\chi(z)\,,
\end{equation}
where $\chi(z)$ is the comoving distance at the same redshift, 
\begin{equation}\label{eq:chi_z}
\chi(z) = \begin{cases}
                      r_c\,\text{sinh}(r/r_c), &\quad  K < 0 \\
                      r, &\quad K = 0 \\
                      r_c\,\text{sin}(r/r_c), &\quad  K > 0\,,
          \end{cases}                    
\end{equation}
with $r_c \equiv 1/(H_0 \sqrt{|\Omega_K|})$. $r$ given by

\begin{align}\label{eq:rz}
    r(z) = \frac{1}{H_0}\int_0^z dz' \frac{1}{E(z')}\,,
\end{align}
with
\begin{eqnarray}\label{eq:ez}
    \nonumber E^2(z)&=&\Omega_r (1+z)^4+\Omega_M (1+z)^3+\Omega_K(1+z)^2\\
    && + \Omega_{\Lambda}(1+z)^{3(1+w_0+w_a)}e^{-3w_az/(1+z)}\,,
\end{eqnarray}
where $\Omega_r,\, \Omega_M,\, \Omega_K,\, \Omega_{\Lambda}$ are the energy density fractions of radiation, matter, curvature and dark energy, respectively, and $w \equiv w_0+w_a(1-a)$ is the time-varying equation of state for dark energy, parameterized by $w_0$, which characterizes the constant part, and $w_a$, which represents the amplitude of time variation~\cite{Chevallier:2000qy,linder:2003}. \refEqs{dL_z}{chi_z} only hold for a supernova that lies in a host galaxy which has no peculiar velocity and observer, in a Friedmann-Robertson-Walker (FRW) universe, such that light from the supernova propagates through a homogeneous and isotropic background. This is not the case for our Universe. 

As light travels from a supernova to an observer, it gets gravitationally lensed by the intervening matter along the l.o.s. In addition, peculiar motion of the host galaxy of the supernova changes the observed redshift of the light. Other effects such as the integrated Sachs-Wolfe effect~\cite{sachs:1967} and gravitational redshifts also affect the observed magnitude and redshift of a given supernova. However, we do not consider them in this paper because these are much weaker than lensing and peculiar velocities~\cite{Hui:2005nm}. At linear order in metric perturbations, the total change in observed luminosity distance can then be written as~\cite{Sasaki:1987ad,futamase/sasaki:1989,Hui:2005nm}
\begin{align}\label{eq:deltad}
\nonumber\delta d_{\text{L}}(z, \hat{\textbf{n}}) = \Bigg[1-\frac{1}{a_s H_s\chi_s}\Bigg]\textbf{v}_{s}\cdot\hat{\textbf{n}}+\frac{1}{a_s H_s\chi_s}\textbf{v}_{o}\cdot\hat{\textbf{n}}&\\
-\frac{3 H^2_0 \Omega_{m0}}{2}\int_{0}^{\chi_s}d\chi \frac{\chi(\chi_s-\chi)}{\chi_s}(1+z)\delta_m(z, \hat{\textbf{n}})&\,.
\end{align}
Here $\hat{\textbf{n}}$ is the unit vector in the \emph{observed} l.o.s. direction, $\chi_s$ is the comoving distance at \emph{observed} redshift $z_s$ of a supernova, $a_s$ is the scale factor corresponding to $z_s$, $H_s$ is the Hubble rate at redshift $z_s$, $\delta_m(z, \hat{\textbf{n}})$ is the matter density fluctuation at redshift $z(\chi)$ in the direction $\hat{\textbf{n}}$, {and $\textbf{v}_{o}$ and $\textbf{v}_{s}$ are the peculiar velocities of the observer and supernova, respectively}. Note that this equation has been derived assuming smallness of metric perturbations but not its derivatives. Therefore, it can still be used to account for some non-linearity in {density and velocity perturbations because they are second- and first- derivatives of the metric perturbation, respectively (assuming a linear relation between density and velocity)}. 

Using \refeq{m_dL_eq} we can write~\cite{Hada:2014jra,Hada:2016dje,Hada:2018ybu}
\begin{align}\label{eq:delm_deld}
\nonumber    \delta m_{\text{obs}}(z, \hat{\textbf{n}}) =\, &5\, \text{log}_{10}(1+\delta d_{\text{L}}(z, \hat{\textbf{n}}))\\
                            \simeq & \frac{5}{\ln\,10}\,\delta d_{\text{L}}(z, \hat{\textbf{n}})\,,
\end{align}
where we have assumed that the fluctuation in luminosity distance is small and linear theory holds. Using \refeq{delm_deld} we can write
\begin{equation}\label{eq:variancem}
    \left \langle \delta m^2_{\text{obs}}(z, \hat{\textbf{n}})\right \rangle = \Big[\frac{5}{\ln\,10}\Big]^2 \left \langle \delta d^2_{\text{L}}(z, \hat{\textbf{n}})\right \rangle
\end{equation}
where the variance of the luminosity distance fluctuation is given by the sum of the variances from lensing and peculiar velocities, 
\begin{align}
    \left \langle \delta d^2_{\text{L}}(z, \hat{\textbf{n}})\right \rangle = \left \langle \delta d^2_{\text{L,lens}}(z, \hat{\textbf{n}})\right \rangle+\left \langle \delta d^2_{\text{L,vel}}(z, \hat{\textbf{n}})\right \rangle\,.
\end{align}
There is no cross-correlation term because the l.o.s. velocities are integrated along the l.o.s. due to the lensing kernel and so average out to zero~\cite{Hui:2005nm}. In addition, we assume that there is no cross correlation between the l.o.s. peculiar velocity of the supernova and the observer. As shown in Ref.~\cite{Hui:2005nm} this contribution is negligible for upcoming surveys. The lensing contribution to the variance is then given as 
\begin{align}\label{eq:lens_var}
    \nonumber \sigma_{\text{lens}}^2(z, \hat{\textbf{n}}) \equiv &\left \langle \delta d^2_{\text{L,lens}}(z, \hat{\textbf{n}})\right \rangle \\
    \nonumber  = & \Big[\frac{3 H^2_0 \Omega_{m0}}{2}\Big]^2\int_{0}^{\chi_s}d\chi \Big[\frac{\chi(\chi_s-\chi)}{\chi_s}\Big]^2\\
    & \times (1+z)^2\int \frac{dk}{2\pi}k P_{\text{nl}}(z, k)\,,
\end{align}
where we have used Limber's approximation and assumed that the redshift bin is not too large. For more details please refer to Appedix D of~\cite{Hui:2005nm}. The velocity contribution is given by
\begin{align}\label{eq:vel_var}
    \nonumber \sigma_{\text{vel}}^2(z, \hat{\textbf{n}}) \equiv & \left \langle \delta d^2_{\text{L,vel}}(z, \hat{\textbf{n}})\right \rangle \\ 
    \nonumber = & \Bigg[1-\frac{1}{a_s H_s\chi_s}\Bigg]^2 \\
    &\times \int \frac{dk}{6\pi^2} [D^{'}(k, z)]^2 P_{\text{nl}}(k, z = 0)\,,
\end{align}
where $P_{\text{nl}}(k,z)$ is the non-linear matter power spectrum at redshift $z$, and $D'(k,z) \equiv -H(z) \frac{d\,D(k,z)}{dz}$ is the growth rate of matter fluctuations. Given the matter power spectrum as a function of redshift, we define the growth factor, $D(k,z)$, as the square root of the ratio of the (non-linear) power spectrum at $z$ to the one at $z=0$, and the growth rate can then be evaluated by numerically differentiation $D$ w.r.t. $z$. 

Note that, in principle, the integrals over the power spectrum in \refeqs{lens_var}{vel_var} range from $0$ to $\infty$. In practice, we apply an exponential cut-off at large $k$ values, $e^{-k^2/k^2_c(z)}$~\cite{Hada:2016dje, Hada:2018ybu}, in order to exclude strong lensing from small scale structures and also because of the uncertainty in modelling the non-linear matter power spectrum on small scales. The cut-off scale $k_c$ is set as follows. First, we define a cut-off mass $M_c$ corresponding to the smallest scale to be excluded, $R_c = (3 M_c/4\pi \rho_{m0})^{1/3}$, where $\rho_{m0}$ is the average matter density at $z=0$. As argued in Ref.~\cite{Hada:2016dje} the lensing efficiency for supernovae at $z\sim 2$ peaks at $z \sim 0.5$, where it is dominated by galaxies. Thus, we choose a mass so as to eliminate lensing from galaxy-size dark matter halos. {Thus, following the discussion in Ref.~\cite{Hada:2016dje}, we set $M_c = 10^{11}M_{\odot}$ which correspond to the typical mass of these halos}. Then, at each redshift $z$, we identify the size of the largest structure that collapses to form a halo. To that end we solve for the largest radius $R(z)$ such that the amplitude of the linear density field smoothed on the scale $R$, $\delta_R(z)$, exceeds the threshold for spherical collapse, $\delta_c/\sqrt{2}$. Then we choose the smaller of $R_c$ and $R(z)$ to define the cut-off scale, $k_c(z) = 2\pi/\text{min}(R_c, R(z))$. This procedure is explained in further detail in Ref.~\cite{Hada:2016dje}. We have also checked that this scale is in fact a conservative choice and integrating the power spectrum down to smaller scales tightens the constraints on neutrino mass as non-linearities on smaller scales are more sensitive to neutrino mass.

Cosmological information contained in the matter power spectrum and the growth rate can help constrain fundamental physics using observations of Type Ia Supernovae alone. In this paper we focus on constraining the sum of neutrino masses and dark energy equation of state using measurements of the luminosity distance of Type Ia Supernovae. 

\subsection{\label{sec:neutrino}Effects of Neutrino Mass on Luminosity Distance}

Massive neutrinos are known to free-stream out of overdense regions and thus suppress growth of structure on small scales, while contributing to growth on larger scales (see e.g. Refs.~\cite{Takada:2005si, Lesgourgues:2006nd}). The exact turn-over point is given by the neutrino free-streaming scale, $k_{\text{fs}}$~\cite{Takada:2005si, Lesgourgues:2006nd}
\begin{equation}\label{eq:kfs}
    k_{\text{fs}}(z) \simeq \frac{0.677}{(1+z)^{1/2}}\frac{m_{\nu}}{1 \text{eV}}(\Omega_{m0})^{1/2} h\,\text{Mpc}^{-1}
\end{equation}
where $m_{\nu}$ is the mass of the $\nu$ neutrino flavour. On wavenumbers $k > k_{\text{fs}}$ the power spectrum is suppressed. Moreover, this suppression increases with time since regions smaller than the free-streaming length continue to grow at a lower rate compared to those that are larger. This results in the growth rate, $D'(k,z)$ becoming sensitive to neutrino mass as well. As a result both the lensing dispersion, \refeq{lens_var}, and the peculiar-velocity dispersion, \refeq{vel_var}, become sensitive to neutrino mass. 

\begin{figure}
    \includegraphics[width = 0.45\textwidth]{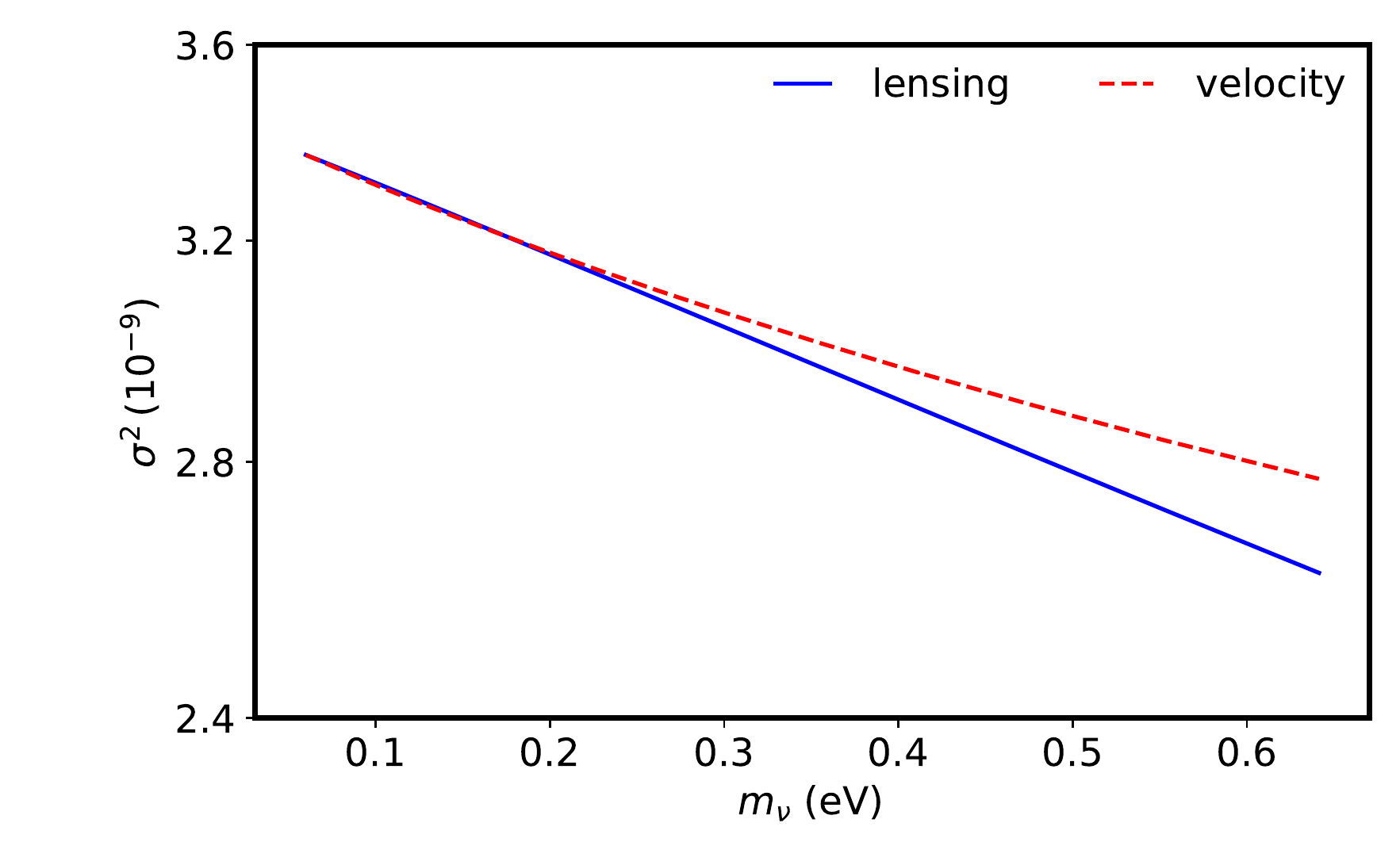}
    \includegraphics[width = 0.45\textwidth]{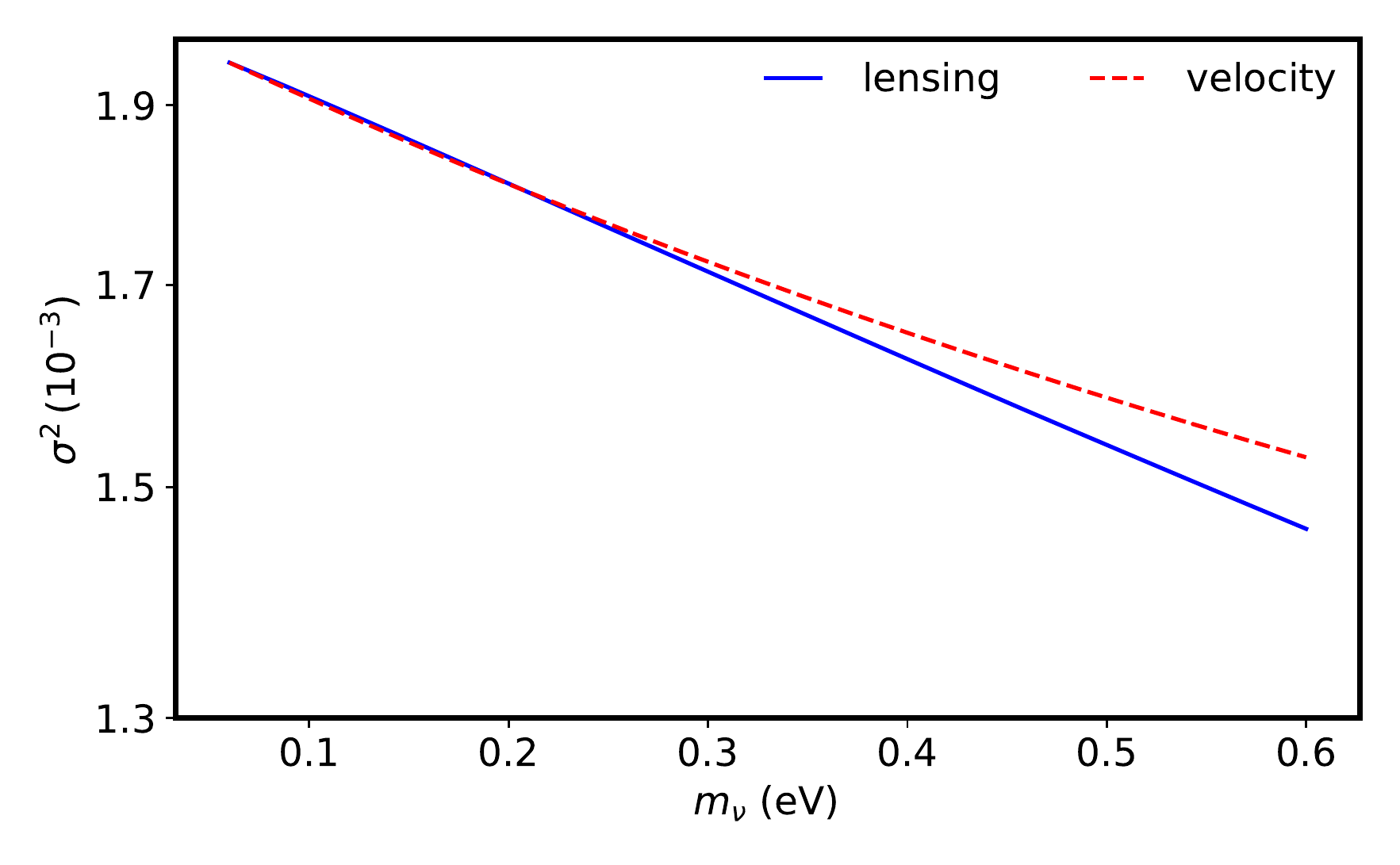}
    \caption{Lensing- (blue solid) and velocity- (red dashed) induced scatter, $\sigma^2$,  in magnitude as a function of the sum of neutrino masses, $m_{\nu}$, in eV for a source at $z= 0.02$ (top) and $z=1$ (bottom). The velocity scatter has been rescaled to have the same amplitude as lensing scatter at $\Sigma m_{\nu} = 0.06$ eV to enable better comparison of slopes. Lensing scatter has a larger slope, and so is more sensitive to the sum of neutrino masses. }
    \label{fig:nu_mass}
\end{figure}

\refFig{nu_mass} shows the effects of neutrino mass on these two dispersions at two different redshifts, $0.02$ (top) and $1$ (bottom). All other cosmological parameters have been kept constant here and we assumed a flat Universe so that $\Omega_{\Lambda} = 1 - \Omega_{\text{tot}}$, where $\Omega_{\text{tot}} = \Omega_r+\Omega_M+\Omega_K$ is the sum of radiation, matter, and curvature densities. We can see that both the lensing and velocity dispersions decrease monotonically with increasing neutrino mass, a signature of the suppressed growth of structure. Moreover, the slope of the lensing dispersion has a larger magnitude than that of velocity at both redshifts, even where peculiar velocity contribution dominates. This indicates that lensing is a more powerful probe of neutrino mass compared to velocities. However, as we show below, the lensing dispersion has a much smaller magnitude at low redshifts and so using peculiar velocities can help in providing information from these redshifts. 

\begin{figure}
    \includegraphics[width = 0.45\textwidth]{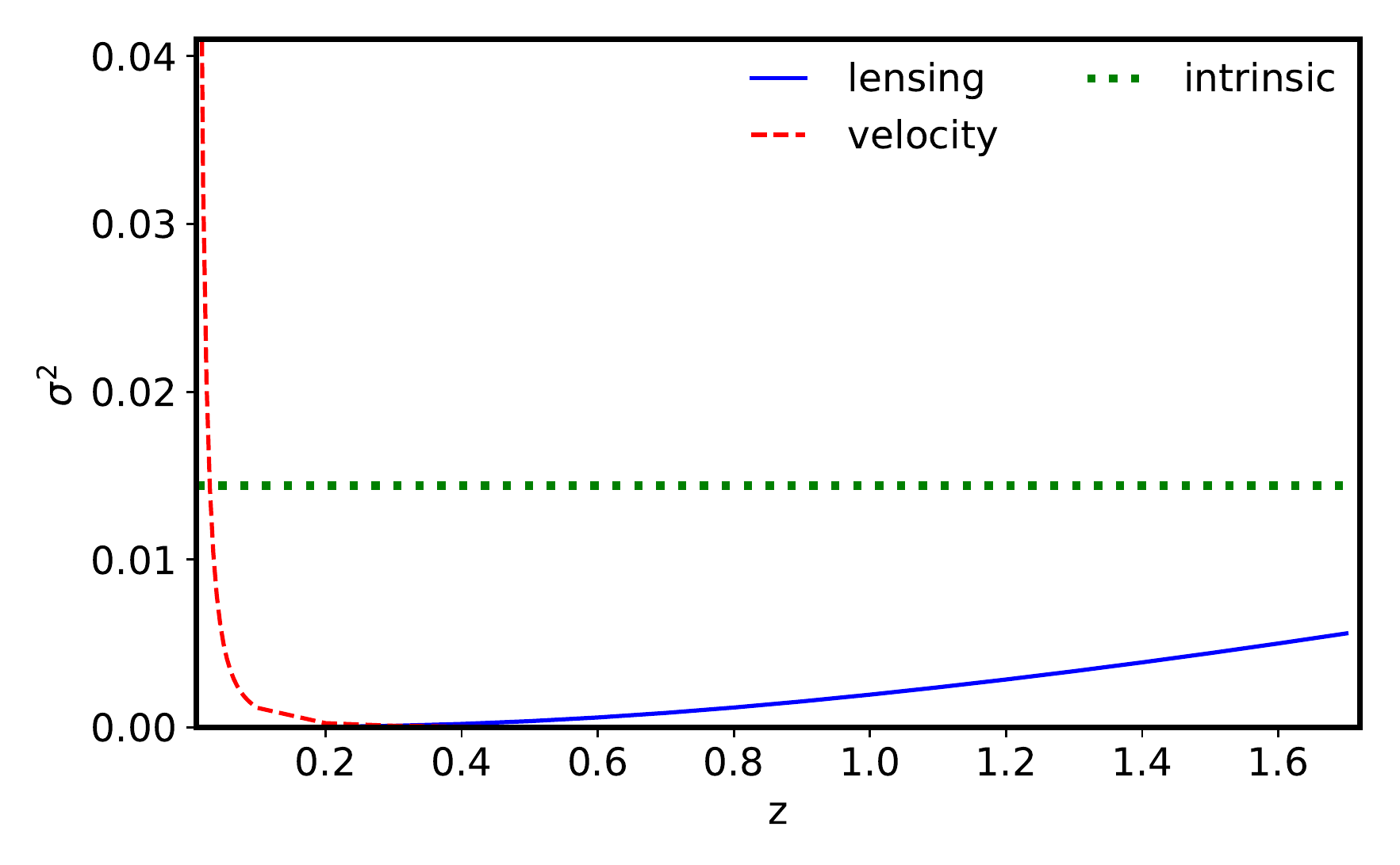}
    \caption{Dependence of magnitude scatter, $\sigma^2$, sourced by lensing (blue solid), peculiar velocities (red dashed), and intrinsic (green dotted) on source redshift ($z$), for $\Sigma m_{\nu} = 0.06$ eV. Lensing produces a large scatter at higher source redshifts while velocities produce a large scatter at lower source redshifts. The intrinsic scatter is independent of source redshift. }
    \label{fig:z_dependence}
\end{figure}

\refFig{z_dependence} shows the expected scatter in magnitude as a function of the source redshift $z_s$ from lensing, peculiar motion, and intrinsic effects for a Type Ia supernova. We assumed the PLANCK $\Lambda$CDM cosmology~\cite{ade:2015} here and assumed a normal hierarchy with $\Sigma m_{\nu} = 0.06$ eV, and 1 massive neutrino and two of them still massless. The non-linear matter power spectrum was calculated using the CLASS code~\cite{Lesgourgues:2011re}, with a non-linear halofit prescription~\cite{Takahashi:2012em}. The intrinsic scatter in the observed magnitude, $\sigma_{\text{int}}$, arising from the intrinsic dispersion in supernova magnitudes, as well as errors due to photometry, light curve fitting and so on is assumed to be $\sigma_{\text{int}} = 0.12$ in accordance with the estimate in Ref.~\cite{Spergel:2015sza}.  We can see that at higher source redshifts the lensing contribution dominates, while at lower redshifts the peculiar velocity contribution is more significant. This is expected {because} at higher redshifts the growth rate is much smaller and the scatter induced by peculiar velocities in \refeq{vel_var} is further suppressed by the increase in $\chi_s$. On the other hand, light travels through more intervening matter when the supernova is at a higher redshift and so the induced deviation from a homogeneous background is larger. In particular, at low redshifts the lensing scatter is even smaller than the intrinsic one and becomes hard to disentangle. 

The redshift dependence of the lensing and peculiar velocity scatter is also useful to isolate it from the intrinsic scatter~\cite{Hada:2018ybu}. From \reffig{z_dependence} it is also clear that a supernova sample covering a wide range of redshifts allows one to constrain cosmology much better than using only low or high redshift supernovae. This will become clearer in \refsec{results} where we demonstrate the effect of adding peculiar velocity information from supernovae at low redshift where we show the constraints on sum of neutrino masses from using both low- and high-$z$ data and from using high-$z$ data alone.  

\section{\label{sec:forecast}Forecasts}
\subsection{\label{sec:survey}Surveys}

\begin{figure}[t]
    \includegraphics[width = 0.47\textwidth]{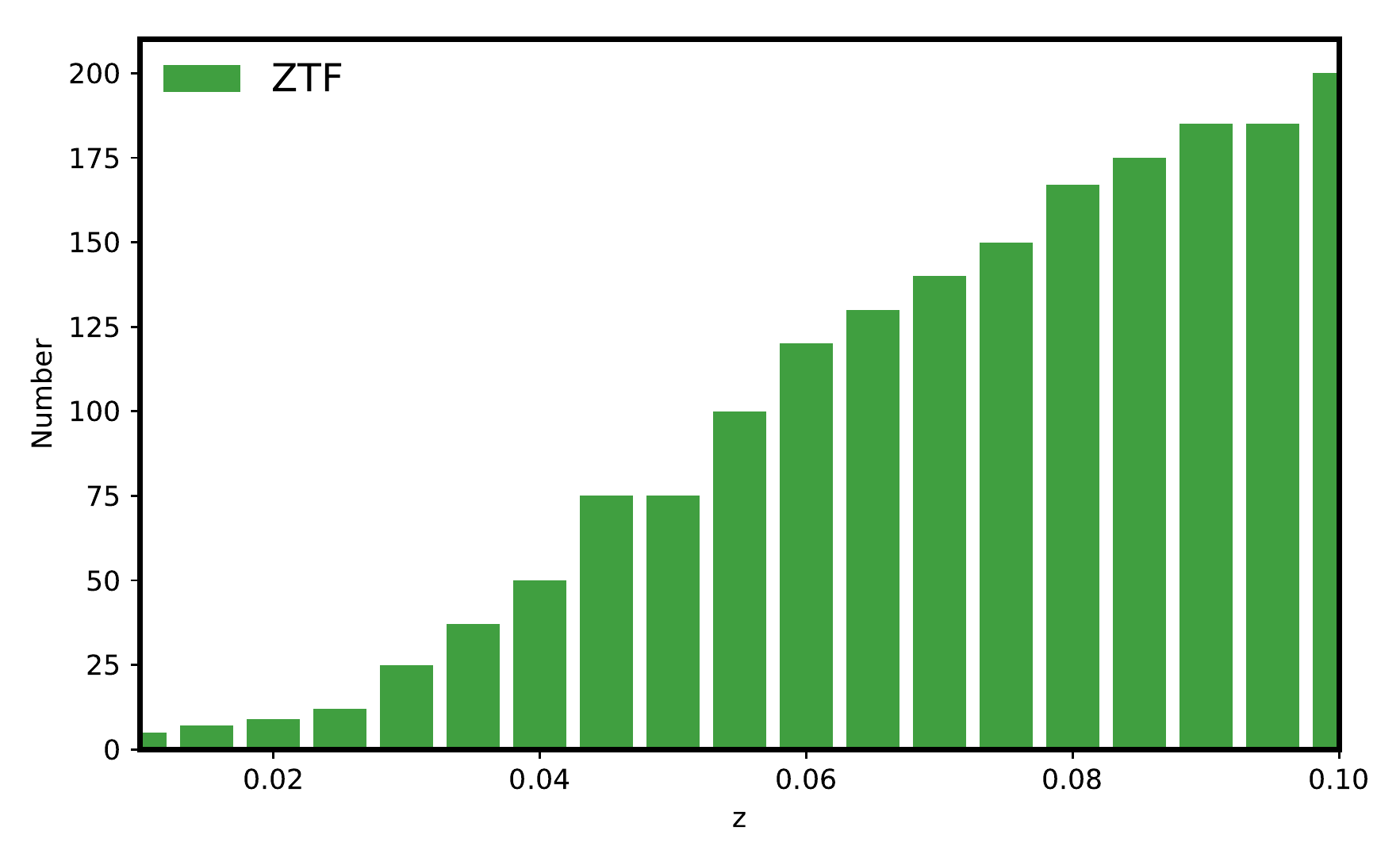}
    \includegraphics[width = 0.47\textwidth]{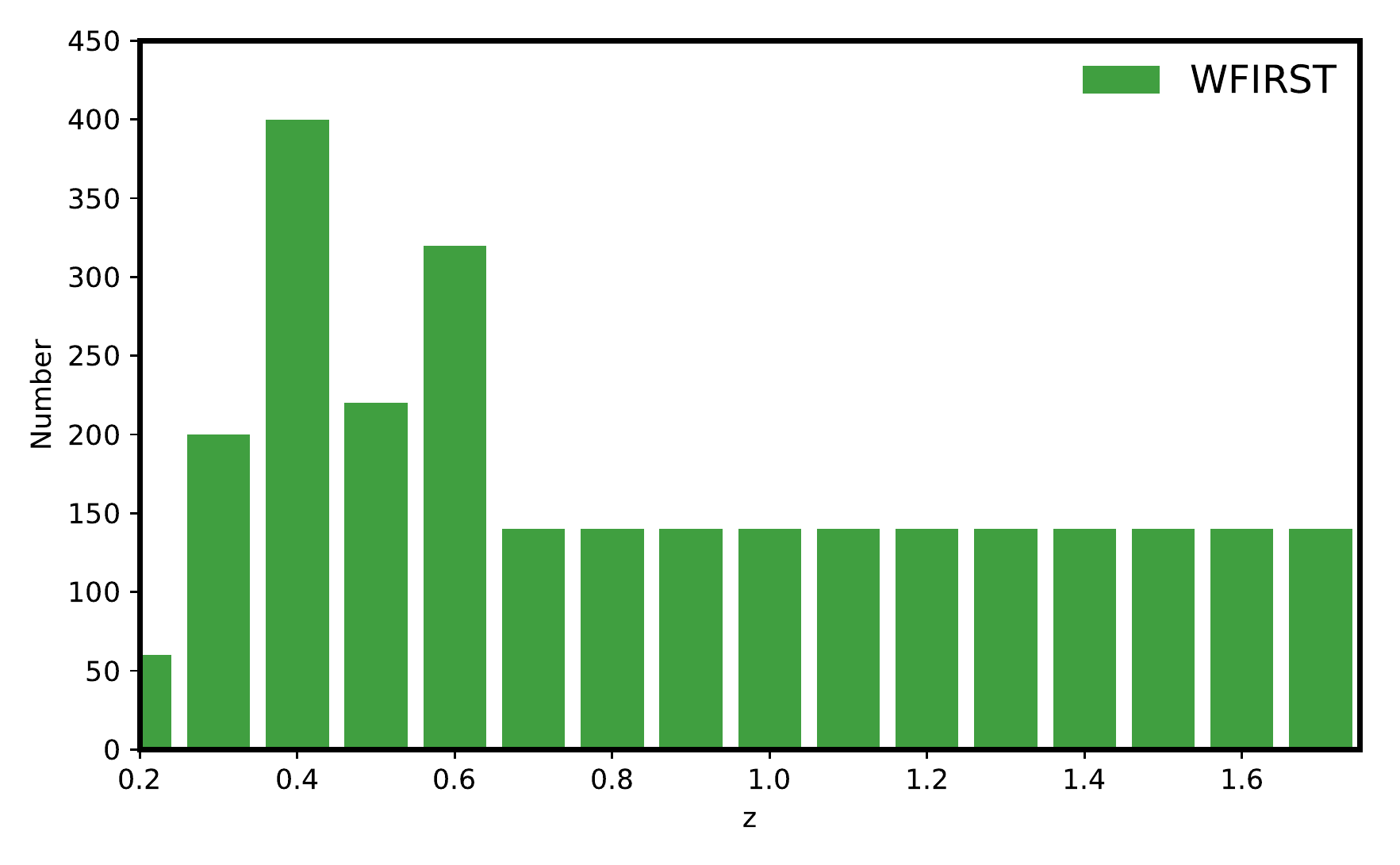}
    \caption{Expected supernova distribution for ZTF~\cite{Graham:2019qsw}(top) and WFIRST~\cite{Spergel:2015sza}(bottom).}
    \label{fig:ztf}
\end{figure}

In \refsec{neutrino} it was shown that a supernova sample spanning a large range of redshifts, from the very low to the very high, is optimal for constraining cosmology. Therefore, to make forecasts for neutrino mass we consider two surveys, the ZTF survey which will observe low redshift supernovae out to $z \lesssim 0.1$~\cite{Graham:2019qsw}, and WFIRST which will observe high redshift supernovae in the range $0.2 \lesssim z \lesssim 1.7$~\cite{Spergel:2015sza}. \refFig{ztf} shows the expected distribution of supernovae in different redshift bins for these surveys. Note that the exact numbers for ZTF were not available, so we estimated the numbers from the distribution given in Ref.~\cite{Graham:2019qsw} and the plot in~\reffig{ztf} shows the distribution used for forecasts. 

\subsection{\label{sec:fisher} Fisher Matrix}
We make use of the Fisher matrix formalism to {obtain} forecasts for neutrino mass and dark energy constraints from the two surveys described in \refsec{survey}. The Fisher matrix is defined as (see for e.g.~\cite{dodelson2003modern})
\begin{align}\label{eq:fisher_def}
    \mathcal{F}_{ab} \equiv -\left \langle \frac{\partial^2 \ln \mathcal{L}}{\partial \theta_{a}\partial \theta_b}\right \rangle\,,
\end{align}
where $\mathcal{L}$ is the likelihood function, and $\theta_i$ represents the $i$-th parameter we want to constrain. The likelihood function gives the probability of finding a particular vector of data $\textbf{d}$ given a vector of parameters $\bm{\theta}$. Thus, the Fisher matrix characterizes how fast the probability of observing $\textbf{d}$ falls off as $\bm{\theta}$ is changed, which is then related to the confidence with which we can estimate $\bm{\theta}$. To evaluate the Fisher matrix we need the likelihood function.

In Refs.~\cite{Hada:2014jra,Hada:2018ybu} a log-normal likelihood has been assumed to describe the lensing-induced scatter in magnitude, motivated by findings that the convergence field is well described by a log-normal distribution~\cite{kayo/taruya/suto:2001,Clerkin:2016kyr}. Additionally, the intrinsic scatter is assumed to be drawn from a Gaussian distribution with a zero mean and variance given by $\sigma^2_{\text{int}}$. Then, the total likelihood is given as a convolution of the log-normal and Gaussian distributions. However, such an approach is hard to implement when we add the contribution from peculiar velocities too. This is because the lensing contribution for a fixed peculiar velocity contribution is not independent of the fixed value assumed for the peculiar velocity contribution. In other words, {$\mathcal{P}(\delta d_{\text{L,lens}}| \delta d_{\text{L,vel}}, \delta d_{\text{L,int}}) \neq \mathcal{P}(\delta d_{\text{L,lens}}| \delta d_{\text{L,int}})$}, where the right hand side (r.h.s.) has been shown to be log-normal distributed~\cite{kayo/taruya/suto:2001}. {Therefore, in this paper we assume that the likelihood function is given by a Gaussian distribution for simplicity. Since a log-normal distribution has two parameters, $\kappa_{\text{min}}$ and $\left\langle\kappa^2\right\rangle$, that depend on cosmological parameters aside from the homogeneous magnitude~\cite{Hada:2018ybu}, constraints on cosmological parameters from using a log-normal likelihood would be tighter than what we obtain here. In other words, our forecast using the Gaussian likelihood would provide more or less conservative constraints. Even if the likelihood were indeed non-Gaussian, we could still expand it around its maximum in a series and the leading order term would be given by a Gaussian}. Also, as pointed out, for example, in~\cite{Adamek:2018rru} if we bin a large enough number of supernovae from different parts of the sky at similar redshifts we can treat the distribution of apparent magnitudes as Gaussian, by the central limit theorem. 

One other assumption we make is that the different supernovae are uncorrelated, so that the total likelihood is just given as a product of the individual likelihoods,
\begin{align}\label{eq:likelihood}
    \ln\mathcal{L}_{\text{tot}} = \ln \prod_{i = 1}^{N} \mathcal{L}_{\text{i}}= \sum_{i = 1}^{N} \ln \mathcal{L}_{\text{i}}\,.
\end{align}
This assumption is motivated by the conclusions of~\cite{Hui:2005nm} which showed that for lensing {the} correlation between different {supernovae } is sub-dominant compared to the individual contribution. For peculiar velocities, though the correlation is as important as the individual contribution, we do not have an exact distribution of the supernovae in the sky and so calculating correlations is not feasible. Hence, our results should be taken as the best constraints possible in absence of the exact survey map. 
\refEq{fisher_def} is linear in $\ln \mathcal{L}$ which implies that from \refeq{likelihood}
\begin{align}
    \mathcal{F}_{ab, \text{tot}} = \sum_{i = 1}^{N} \mathcal{F}_{ab, \text{i}}\,,
\end{align}
which is simply a sum of matrices. The constraints on parameters are then obtained by inverting the Fisher matrix, $\mathcal{F}_{ab, \text{tot}}$. 

With these assumptions, we can write the log-likelihood for a single supernova observed to be at redshift $z$ and in direction $\hat{\textbf{n}}$ as
\begin{align}\label{eq:likelihood_1}
    \nonumber\ln \mathcal{L} = &-\frac{(m_{\text{obs}}(z, \hat{\textbf{n}})-m_{\text{homo}}(\bm{\theta}, z, \hat{\textbf{n}}))^2}{2\,\sigma^2_{\text{tot}}(\bm{\theta}, z, \hat{\textbf{n}})}\\
    &-\frac{1}{2}\ln \sigma^2_{\text{tot}}(\bm{\theta}, z, \hat{\textbf{n}})\,,
\end{align}
where we have explicitly indicated the terms that depend on cosmological parameters $\bm{\theta}$, $m_{\text{homo}}$ is the magnitude of the supernova that would have been observed in the absence of inhomogeneities, and $\sigma^2_{\text{tot}} \equiv \sigma^2_{\text{lens}}+\sigma^2_{\text{vel}}+\sigma^2_{\text{int}}$ is the total variance of the difference in observed and homogeneous magnitudes. The average in \refeq{fisher_def} is to be taken over $m_{\text{obs}}$, where $\left \langle m_{\text{obs}} \right \rangle \equiv m_{\text{homo}}$. As a result, any terms linear in $m_{\text{obs}}-m_{\text{homo}}(\bm{\theta})$ in the second derivative of the log-likelihood (see \refeq{fisher_def}) average out to $0$. Carrying out the {second} derivatives in \refeq{fisher_def}, for a supernova at $z_i$, we get
\begin{align}\label{eq:single_l}
    \nonumber \mathcal{F}_{ab, i} = &\Bigg[\frac{1}{\sigma^2_{\text{tot}}}\frac{\partial m_{\text{homo}}(z_i)}{\partial \theta_a}\frac{\partial m_{\text{homo}}(z_i)}{\partial \theta_b}\\
    &+\frac{1}{2\,\sigma^4_{\text{tot}}}\frac{\partial \sigma^2_{\text{tot}}(z_i)}{\partial \theta_a}\frac{\partial \sigma^2_{\text{tot}}(z_i)}{\partial \theta_b}\Bigg]\,,
\end{align}
from which the total likelihood of a supernova sample distributed in redshift is given as
\begin{align}\label{eq:tot_l}
    \nonumber \mathcal{F}_{ab, \text{tot}} = \sum_{i=1}^{N} &\Bigg[\frac{1}{\sigma^2_{\text{tot}}}\frac{\partial m_{\text{homo}}(z_i)}{\partial \theta_a}\frac{\partial m_{\text{homo}}(z_i)}{\partial \theta_b}\\
    &+\frac{1}{2\,\sigma^4_{\text{tot}}}\frac{\partial \sigma^2_{\text{tot}}(z_i)}{\partial \theta_a}\frac{\partial \sigma^2_{\text{tot}}(z_i)}{\partial \theta_b}\Bigg]\,.
\end{align}
The likelihood is simply given by first derivatives of the magnitude in a homogeneous universe and the variances given by \refeqs{lens_var}{vel_var}. These derivatives are easily calculated using the central difference formula numerically, and can then be used to build up the Fisher matrix. 

\begin{figure}
    \includegraphics[width = 0.51\textwidth]{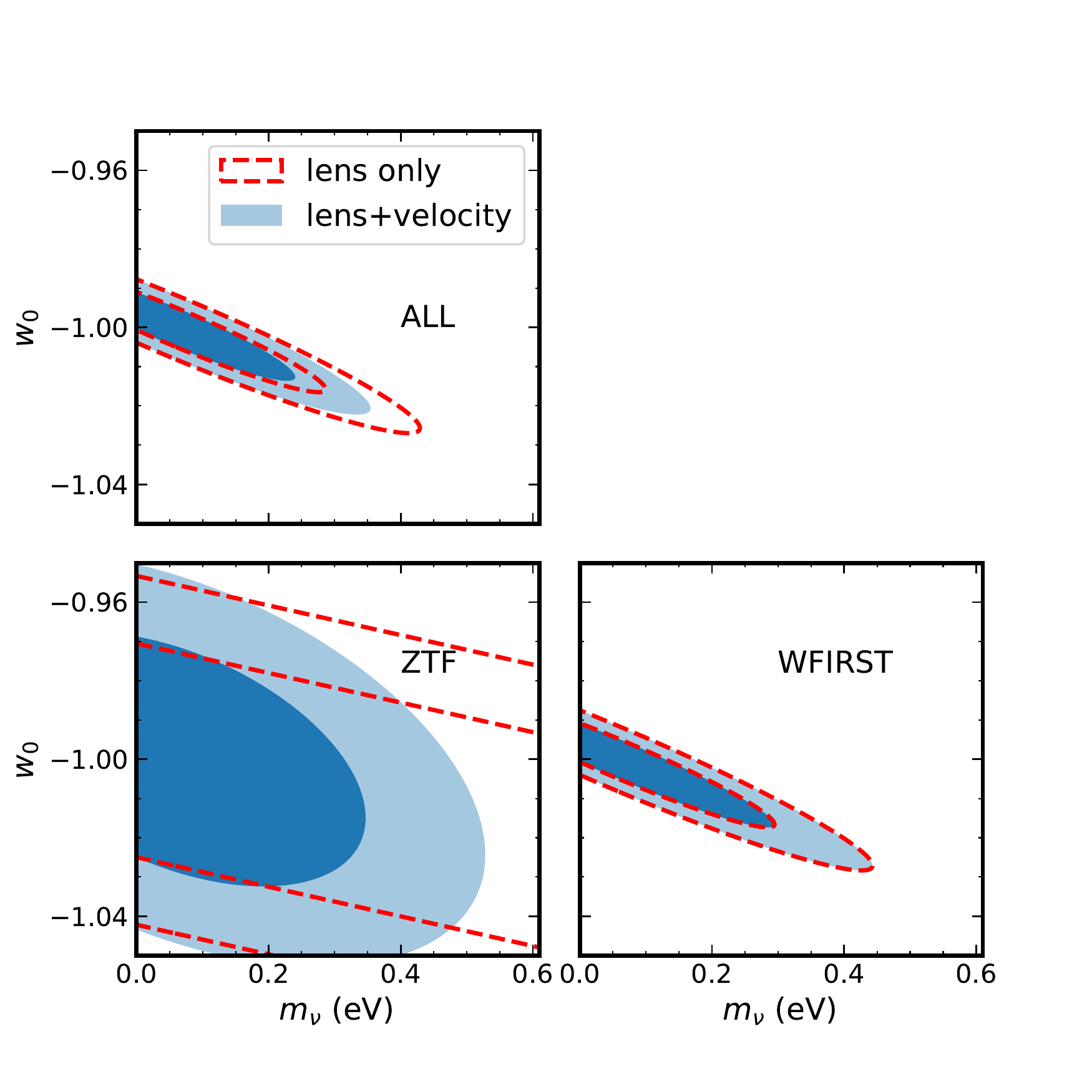}
    \caption{Expected constraints on $w_{0}$  and sum of neutrino masses $m_{\nu}$ in eV from the full supernova {samples of ZTF and WFIRST (top left), ZTF alone (bottom left) and WFIRST alone (bottom right)}. 
    Red dashed curves show the $1-$ and $2-\sigma$ constraints when using lensing scatter alone. Blue filled regions denote the same when using both lensing and peculiar velocities. Note that all other cosmological parameters have been fixed to their fiducial values for these constraints. At low redshifts (bottom left) peculiar velocities are crucial to derive any meaningful constraints on the sum of neutrino masses from supernovae alone.}
    \label{fig:mnu_w}
\end{figure}

\subsection{\label{sec:results} Results}
The first parameter combination we consider is $\Sigma m_{\nu}$ - $w_0$. \refFig{mnu_w} shows the 1- and 2-$\sigma$ contour plots obtained using the full supernova sample from ZTF and WFIRST as well as from using either of the two surveys, and keeping all other cosmological parameters fixed. The dashed red contours show the 1- and 2-$\sigma$ {confident levels} obtained when considering scatter due to lensing alone, while the filled blue contours show the effect of adding information from scatter due to peculiar velocities. Neutrino mass contributes to the Fisher information, \refeq{single_l}, through the derivative of the magnitude in a homogeneous Universe as well as through the dependence of the matter power spectrum via the second term of {the equation}. The parameter $w_0$ also affects the Fisher matrix both through the homogeneous magnitude, where it is anti-correlated with neutrino mass (c.f.~\refeq{ez}), and through the magnitude scatter, where a large neutrino mass and small $w_0$ serve to suppress structure formation. Thus the overall slope of this combination is negative. Adding peculiar velocities only from $z < 0.1$ improves the constraint on neutrino mass by about $0.04$ eV, which is almost a $15 $\% improvement over the case without peculiar velocities. This represents our best constraints.  

In the bottom right panel of~\reffig{mnu_w} we see that when using only high redshift supernova data adding peculiar velocities does not help at all. This is expected because as we showed in \refsec{neutrino} the scatter from peculiar velocities is a sharply decreasing function of redshift and so becomes negligible when we include only the WFIRST sample which is dominated by supernovae at $z > 0.4$. 

In contrast, the bottom left panel of~\reffig{mnu_w} shows what happens when we restrict ourselves to low redshift information alone. Here, as expected, the lensing contribution hardly constrains the neutrino mass at all. Using peculiar velocity information {however,} we find that neutrino mass can still be constrained to $\lesssim 0.3$ eV at $1-\sigma$ level. In fact it is crucial in this case to include peculiar velocity information, otherwise there is almost no constraining power in lensing. Note also that these constraints come from using less than $2000$ supernovae from the ZTF sample. By combining with the low-$z$ data from LSST for example, these constraints can be improved further. 

\begin{figure}
    \includegraphics[width = 0.5\textwidth]{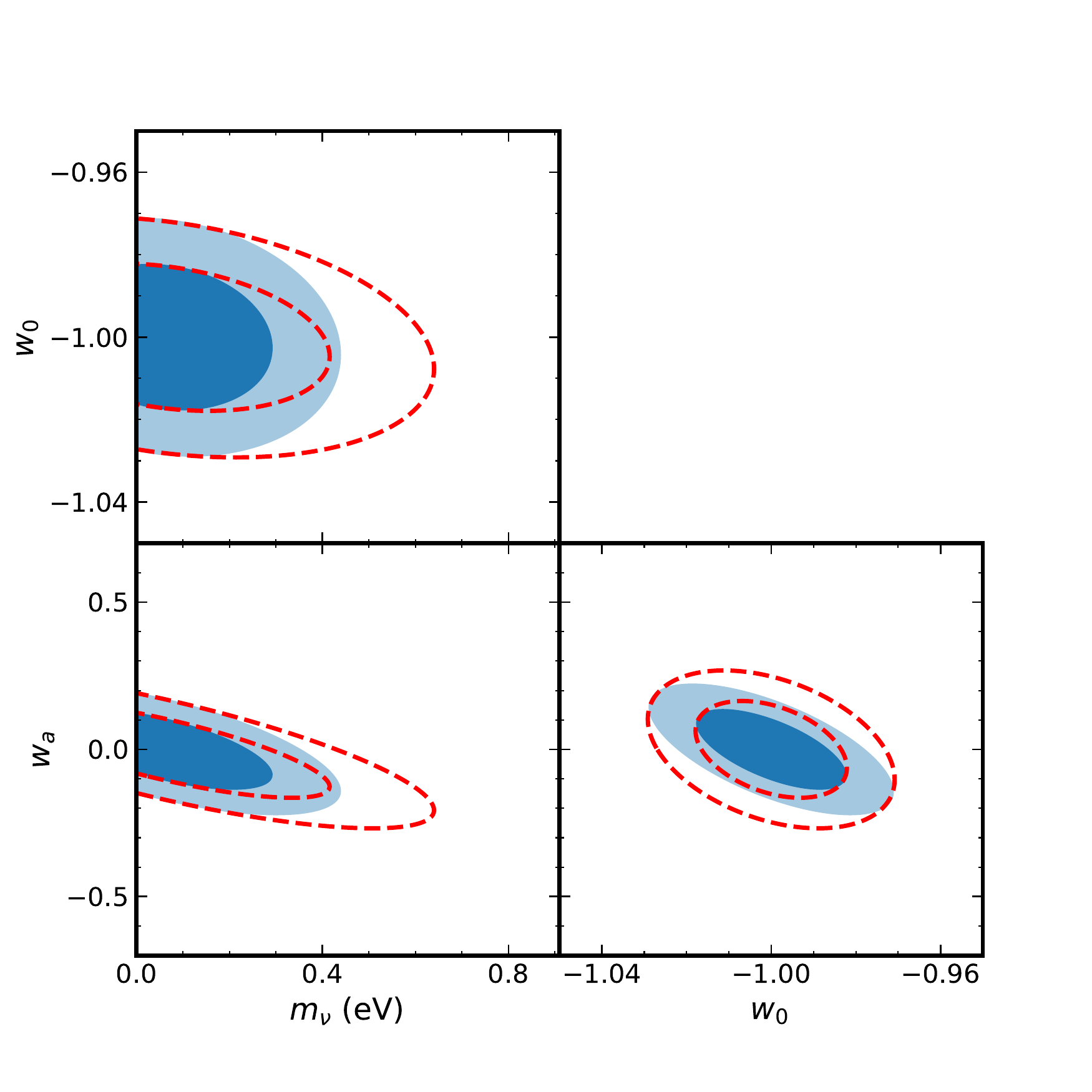}
    \caption{Joint constraints on $\Sigma m_{\nu}$, $w_0$, and $w_a$ with all other cosmological parameters fixed, using the full supernova samples of ZTF and WFIRST. As before, red dashed lines show constraints when using lensing alone, while blue filled regions show constraints when adding peculiar velocities. Note also that varying $w_a$ weakens the neutrino mass constraint, as evident from the larger range of $m_{\nu}$ (eV) compared to~\reffig{mnu_w}. There is no improvement in $w_0$ but $w_a$ and $m_{\nu}$ are still better constrained when peculiar velocities are added.}
    \label{fig:wwm}
\end{figure}

Next we consider switching on the time variation of dark energy, $w_a \neq 0$. The constraints obtained when all three are varied are shown in~\reffig{wwm}. Each panel shows joint constraints on two parameters and the third parameter has been {marginalized} over. All other parameters are fixed to their fiducial values. The first striking feature here is that once $w_a$ is allowed to vary, adding velocities does not help at all in constraining $w_0$~\cite{Linder:2003dr}. Moreover, when we marginalize over the sum of neutrino masses, we see a significant improvement in constraining $w_a$ on adding peculiar velocities. Similarly, adding velocities when marginalizing over $w_0$ or $w_a$ leads to improvement in neutrino mass constraints. The impact of adding peculiar velocity information can be quantified using the volume of the parameter ellipsoid which $\propto 1/\sqrt{\det{\mathcal{F}_{ab}}}$. For these three parameters, and the two surveys we consider, {$\det\mathcal{F}_{ab,\text{lensing}} = 5.54 \times 10^6$
whereas $\det\mathcal{F}_{ab,\text{lensing}+\text{velocity}} = 1.24\times 10^7$}
which implies that adding velocities shrinks the volume by $\sim 33$\%.

\begin{figure}
    \includegraphics[width = 0.5\textwidth]{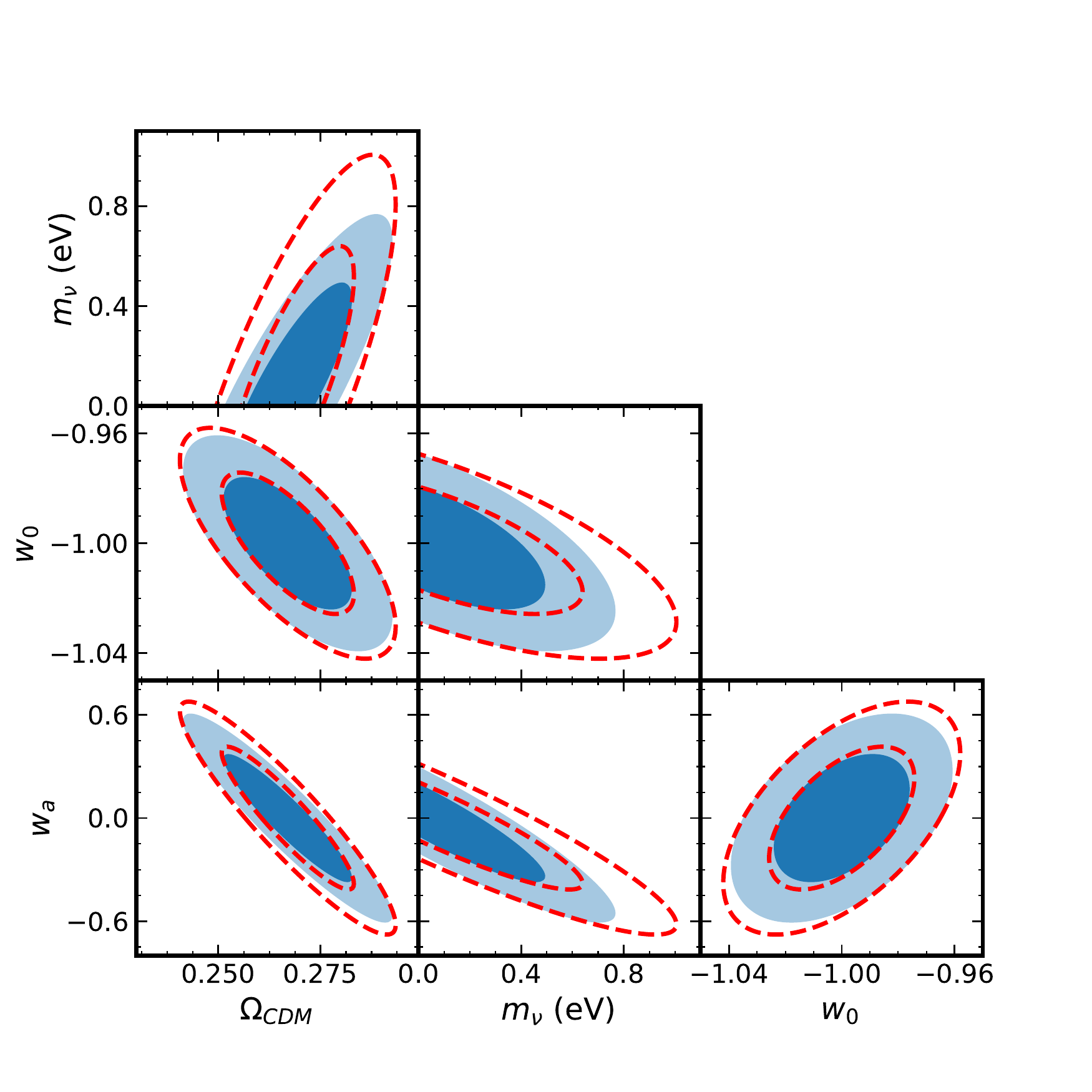}
    \caption{Joint constraints on $\Omega_{\text{CDM}}$, $\Sigma m_{\nu}$, $w_0$, and $w_a$ with all other cosmological parameters fixed. As before, red dashed lines show constraints when using lensing alone, while blue filled regions show constraints when adding peculiar velocities. Varying $\Omega_{\text{CDM}}$ simultaneously degrades constraints by a factor of $\sim 2$, and once again we scale the $m_{\nu}$ axis to account for the weaker constraint.}
    \label{fig:wwom}
\end{figure}

Finally, in~\reffig{wwom} we show the constraints obtained when we also vary $\Omega_{\text{CDM}}$. As expected, the constraints degrade, but we are still able to constrain $-0.7 \lesssim w_a \lesssim 0.7$, which is quite encouraging. However, now the sum of neutrino masses that supernovae alone can constrain becomes much larger, $\lesssim 1$ eV. In general we find that the constraints weaken by a factor $\sim 2$ if $\Omega_{\text{CDM}}$ is not well constrained. 

Our results show that neutrino mass constraints \emph{always} improve when adding peculiar velocity information from supernovae. Similar results have also been obtained for galaxy redshift surveys~\cite{Boyle:2017lzt}. While not quite as competitive as the constraints obtained with galaxy redshift surveys, it is still interesting to note that we can constrain neutrino mass and dark energy equation of state to that accuracy using $\sim 4000$ supernovae alone.  

\section{\label{sec:conclusion}Conclusion}
{We} have calculated the expected constraints on {the} sum of neutrino masses and equation of state of dark energy that can be obtained using information from the lensing and peculiar velocities of supernovae. Standard cosmological analyses from supernovae assume that light propagates in a homogeneous Universe. While this assumption holds good for low-redshift supernovae, it is no longer true as we observe supernovae at higher redshifts. Intervening matter along the l.o.s. lenses the supernovae and makes them appear brighter or fainter, thus leading to a deviation from the $\Lambda$CDM prediction. At low redshifts, we also showed that peculiar velocities contaminate the magnitude quite significantly, and can be modelled to extract neutrino mass information. 

We derived the Fisher matrix for the observed magnitude when it is given by a sum of the homogeneous magnitude and corrections due to lensing and peculiar velocities, and used it to make forecasts on the sum of neutrino masses and dark energy equation of state for two future surveys, the ZTF which is a low-redshift survey, and WFIRST which is a high-redshift survey. Our results show that using {data of about $4000$ supernovae} out to $z \sim 1.7$ from only these two surveys can help constrain $\Sigma m_{\nu} \lesssim 0.31$ eV, $\sigma(w_0) \lesssim 0.02$, and $w_a \lesssim 0.18$ if all other parameters are fixed. We also showed that peculiar velocity information is crucial to constraining the sum of neutrino masses if we allow other cosmological parameters to vary, or if we only focus on low-redshift supernovae. When allowing a time-varying equation of state for dark energy, we showed that peculiar velocities can allow significant improvements in constraining $w_a$ if the sum of neutrino masses is marginalised over. Interestingly, we do not see improvement in constraints on $w_0$ once $w_a$ is allowed to vary. Overall, adding peculiar velocities provides $\sim 33$\% reduction in volume in the $\Sigma m_{\nu}$-$w_0$-$w_a$ parameter space. Future surveys such as LSST will measure an even larger number of supernovae, covering both low and high redshifts, which will considerably shrink the error bars on neutrino mass and dark energy equation of state, making supernovae a competitive complementary probe to galaxy redshift surveys. 

\begin{acknowledgments}
We would like to thank Ryuichiro Hada for his patient help in answering questions on his papers and Masato Shirasaki for helpful discussions.
T.~O. acknowledges support from the Ministry of Science and Technology of Taiwan under Grants No. MOST 106-2119-M-001-031-MY3 and the Career Development Award, Academia Sinica (AS-CDA-108-M02) for the period of 2019 to 2023. 
T.~F. is supported by Grant-in-Aids for Scientific Research from JSPS (No. 17K05453 and No. 18H04357).
\end{acknowledgments}

\bibliography{references}

\end{document}